\documentclass[noshowpacs, twocolumn,preprintnumbers,secnumarabic,amsmath,amssymb,nofootinbib]{revtex4-1}
\usepackage{dcolumn}      
\usepackage{bm}           
\usepackage{graphicx}
\usepackage[utf8]{inputenc}
\usepackage[spanish]{babel}
\usepackage{csquotes}
\usepackage{natbib}
\usepackage{caption}
\usepackage{wrapfig}
\usepackage{gensymb}
\usepackage{float}
\usepackage{subcaption}
\usepackage{amsmath,amssymb}  
\usepackage{bm}  
\usepackage[font=small,labelfont=bf]{caption}

\usepackage[pdftex,bookmarks,colorlinks,breaklinks]{hyperref}  
\usepackage{mathrsfs}
\usepackage{amsmath}

\begin{document}
\title{Generalised Maxwellian Equations and Implications}
\author{Bob Osano} 
\email{bob.osano@uct.ac.za} \affiliation{Astrophysics, Cosmology and Gravity Centre, Department of Mathematics and Applied Mathematics, University of Cape Town, Rondebosch 7701, Cape Town, South Africa\\\\}
\affiliation{Academic Development Programme, Science, Centre for Higher Education Development, University of Cape Town, Rondebosch 7701, Cape Town, South Africa}
\begin{abstract}
We present a reformulation of Mawxell's equation and examine the consequence of this new formulation. We argue that studies of such diverse topics as magnetic monopole, magnetic reconnection and magneto-genesis could benefit from the new formulation.
\end{abstract}
\date{\today}
\maketitle
\section{Introduction}
The starting point of what we call Maxwell's equations can be traced back to Thomson who worked on application of analogies. Maxwell's article in 1856 that dealt with analogues of hydrodynamics with an incompressible and massless fluid is arguably the starting point of the development of the equations that bear his name. That article catalogs two crucial steps that eventually led to the unification of electricity and magnetism; 1) a classification of quantities into two kinds based on whether the connoted quantity or intensity is suggested and, 2)  a demonstration of the existence of a linear relationship between the parameter is given. These are crucial because they were a foundation for the development that followed. Magnetic intensity was denoted by ${\bold H}$, the electric intensity by ${\bold E}$ Similarly, the magnetic quantity is denoted by ${\bold B}$, the electric quantity by ${\bold D}$ respectively. The coefficients $\mu$ and $\epsilon$ represent the ling between magnetic quantity and intensity, and electric quantity and intensity respectively. On applying Stokes's and Gauss's integral theorems, Maxwell was able to express all known basic electromagnetic laws in terms of differential divergence and curl operations (in component form), each relating to some analogy in hydrodynamics. He interpreted the magnetic intensity {\bf H} as the angular velocity of a vortex of physical fluid whose mass corresponded to the magnetic permeability. 

Maxwell found that he could imitate Ampere's and Faraday's laws in curl equation form, upon separating the vortex tubes. The vortex tubes corresponded to electric charges and could be taken as the conveyors of the rotating motion between vortices. But this did explain the electrostatic polarisation,  and so Maxwell decided to add elasticity to the fluid vortices. This addition corresponded to a displacement current term in his equations. The result was that motions were no longer felt simultaneously but they propagated through the structure in a wavelike manner. The velocity of the resulting electromagnetic wave was close to the velocity of light, confirming what experiments had shown. This finally unified electro-magnetics and light. The concept of displacement-current remained a prediction of this theory awaiting the design of high-frequency equipment, which only came in the beginning of the 20th century. The concept of displacement current has generated often conflicting views in literature as is demonstrated in \cite{Roche,Jack, Roche2}. We are however interested in the methodology and physical relevance  and will not delve into this debate.

Maxwell's original formulation involved 20 scalar equations with as many unknowns and it was Heaviside who regrouped them into the set we call Maxwell's equations. In particular, he removed the unnecessary potential quantities and was left with only the field and source quantities in a symmetric form.

In this context, Heaviside began to develop vector algebra as what he considered to be the proper mathematical language for the discussion of 3-D force fields \cite{Land}. He was convinced that the formulation should be in terms of fields (${\bold E}$ and ${\bold H}$) rather than potentials as Maxwell had them, and in the general vector notation which made the inherent symmetry and the refinement of the equations clear. Maxwell's ( and by extension Heaviside) restricted himself to very special cases, but the method that he developed allows for greater generalisation with potentially for greater applications. This is what we will examine in this article. 
\section{Modification of Maxwell's equations}
There are at least two methods one can follow in order to give a consistent modification of Maxwellian equations:

1) The first approach is to attempt qualitative and quantitative changes to the basic quantities and then re-derive the full set equations by going through the modifications of Faraday's and Ampere's laws. One could then add elasticity to the fluid vortices to obtain a term that corresponds a displacement current term. Examples of where this approach has been used in \cite{Pinh,Henz} where a modification of Maxwell's equations in vacuum is done. The same approach was also used in \cite{Ger}, where the author demonstrates that Maxwell equations can be derived from first principles, in a method similar to that which has been used to derive the Dirac relativistic electron equation. Another example where this approach features is in \cite{WalkB, JohnP}, where a reinterpretation of Maxwell's equation is given and the inclusion of the Lorentz force is in Maxwell's equations is explored. The authors argue that the new formulation is consistent with Maxwell's equations for bodies at rest and a new kind of electromotive force is identified.

2) A second approach is to assume that the modification has to obey the same symmetries as those of classical Maxwell's equations. Such an approach is considered in \cite{Paris}. Here we begin by assuming the existence of relation between the fundamental variables. We will adopt this approach in the rest of the article.

\section{New variables}
Assume the parameter  ${\bold E}$, ${\bold D}$,   ${\bold H}$,  ${\bold B}$ and ${\bold J}$ are the electric field, electric displacement, magnetic field, magnetic flux and  current density respectively. Maxwell's equations are linkages between these coupled parameters. In particular $ \bold{D}=\bold{D}(\bold{E}, \bold{H})$ and $\bold{B}=\bold{B}(\bold{E}, \bold{H}).$
The electric properties are a lot more complicated, given the fact that they may depend on microscopic properties of the media on the one hand, and macroscopic properties such as density and temperature on the other. In general one can approximate the relationships as follows: \\
\[
{\bold D}={\bold E}+4\pi{\bold N}, ~~~~{\bold B}={\bold H}-4\pi{\bold M},
\]
where  $\bold{N}$ is the electric polarisation vector and $\bold{M}$ the magnetisation. Note that if both the polarisation and the magnetisation are ignored, one recovers the standard parameters; ${\bold D}={\bold E}$ and ${\bold B}={\bold H}$. On the other hand if only ferro-electric and ferro-magnetic effects are ignored and the fields are relatively small, then one could define a new time and spatial dependent magnetic parameter by ${\bold{\mathcal{B}}}=\alpha {\bold B}$ and the corresponding electric parameter be given by ${\mathcal{E}}=\beta{\bold E}$. Note that the current can also be written in the form $\mathcal{J} =\gamma {\bold J}$.  In special cases, $\alpha$ and $\beta$ maybe take as dielectric and permeability tensors respectively, which reduce to scalar functions in the case of {\it isotropic mediums}. To be clear these scalars encode effects that may be due to media properties and environmental dynamics and as a consequence should not be viewed as simply dielectric and permeability respectively. It is this scope of what else they might encode that is the core subject of this article. 
 \section{Maxwell's equations}
One can easily show that the corresponding set of Maxwellian equations for the general {\it anisotropic} case take the form
\begin{eqnarray}
\label{eq1}\dot{\mathcal{B}}&=&-\frac{\alpha}{\beta}\nabla\times{\mathcal{E}}+( \frac{\dot{\alpha}}{\alpha}{\mathcal{B}}+\frac{\alpha}{\beta^2}{\mathcal{E}}{.}\nabla\beta)\\
\label{eq2}\dot{\mathcal{E}}&=&\frac{\beta}{\alpha}\nabla\times{{\mathcal{B}}-\frac{1}{\gamma}\mathcal{J}}+( \frac{\dot{\beta}}{\beta}{\mathcal{E}}-\frac{\beta}{\alpha^2}{\mathcal{B}}{.}\nabla\alpha)
\end{eqnarray}
while the constraint equations take the form,
 \begin{eqnarray}
\label{eq3} \nabla {.}{\mathcal{B}}&=&\frac{{\mathcal{B}}{.}\nabla\alpha}{\alpha}\\
\label{eq4} \nabla{.}{\mathcal{E}}&=&\beta\rho_{0}+\frac{{\mathcal{E}}{.}\nabla\beta}{\beta},
 \end{eqnarray} where $\rho$ is the material density. We have adopted the convention $c^2=1$  and $\mu_{0}=1$
The analysis of these coupled equations will generally yield the full spectrum of electromagnetic effects. It is important to note that not all predictions from these sets of equations have physically verifiable effects and restrictions are needed in order to recover the known results. We need to put a caveat in the above statement, in particular that although some predictions from the theory may not have been observed, this should not be taken to mean that such results are unattainable. In the following sections, we make the assumption that both scalars obey the same conditions with respect to time and spatial variations. We now turn to the sub cases.

\subsection{\label{ST}Constant scalars with respect to both time and spatial variations}
Consider the case where both scalars are constant with respect to time and distance: \[\dot{\alpha}=0=\nabla\alpha, ~~\dot{\beta}=0=\nabla\beta.\]  It is evident that the bracketed terms in Eqs. (\ref{eq1} and \ref{eq2} ), likewise the last terms of the RHS of Eqs (\ref{eq3}) and (\ref{eq4}) vanishing. This recovers Maxwell's evolution equations: 
\begin{eqnarray}
\label{eq5}\dot{\mathcal{B}}&=&-\frac{\alpha}{\beta}\nabla\times{\mathcal{E}}\\
\label{eq6}\dot{\mathcal{E}}&=&\frac{\beta}{\alpha}\nabla\times\mathcal{B}-\frac{1}{\gamma}{\mathcal{J}}
\end{eqnarray}
and constraint equations
\begin{eqnarray}
\label{eq7} \nabla {.}{\mathcal{B}}&=&0\\
\label{eq8} \nabla{.}{\mathcal{E}}&=&\beta\rho_{0}.
\end{eqnarray} It is then possible to develop the equation taking into account the usual considerations of Ohms or Ampere's laws. This kind of development can be found in literature and will not be repeated in this short article. What we would like to do is to consider the consequence of non constant $\alpha$ or $\beta$, and what kind of theory this might potent.
 \subsection{Constant with respect to time variations}
In this section we consider the sub case involving $\alpha$ and $\beta$ being constant with respect to time, but not in space. In particular, we require $\dot{\alpha}=0=\dot{\beta}.$ 
In this regard Eqs. (\ref{eq1}-\ref{eq2}) take the form:
\begin{eqnarray}
\label{eq13}\dot{\mathcal{B}}&=&-\frac{\alpha}{\beta}\nabla\times{\mathcal{E}}+(\frac{\alpha}{\beta^2}{\mathcal{E}}{.}\nabla\beta)\\
\label{eq14}\dot{\mathcal{E}}&=&\frac{\beta}{\alpha}\nabla\times{{\mathcal{B}}-\frac{1}{\gamma}\mathcal{J}}-(\frac{\beta}{\alpha^2}{\mathcal{B}}{.}\nabla\alpha)
\end{eqnarray}
while the constraint equations take the form:
 \begin{eqnarray}
\label{eq15} \nabla {.}{\mathcal{B}}&=&\frac{{\mathcal{B}}{.}\nabla\alpha}{\alpha}\\
\label{eq16} \nabla{.}{\mathcal{E}}&=&\beta\rho_{0}+\frac{{\mathcal{E}}{.}\nabla\beta}{\beta}.
 \end{eqnarray}
 Eq (\ref{eq15}) suggests that the spatial dependence of $\alpha$ leads to a monopole \cite{Dir,GOD, Roh} and hence a possible pathology in this theory unless we are willing to live with this or accept that no pathology exists and that physical predictions are experimentally verifiable. Note the term on the RHS of Eq (\ref{eq15}) may be taken as a magnetic charge density, while the last term on the RHS of Eq(\ref{eq13}) the magnetic current. These terms used to be dropped in the belief that magnetic monopoles have not been observed, unless the results of \cite{Mono} are taken as a confirmation of their existence. Whichever bias one holds, it remains that the present formulation lends itself to the study of magnetic monopoles.
 
\subsection{\label{Max} Constant with respect to spatial variations}
In this section we consider the sub case involving $\alpha$ and $\beta$ being constant in space, but not in time. In this case Eqs. (\ref{eq1}-\ref{eq2}) take the form:
\begin{eqnarray}
\label{eq9}\dot{\mathcal{B}}&=&-\frac{\alpha}{\beta}\nabla\times{\mathcal{E}}+( \frac{\dot{\alpha}}{\alpha}{\mathcal{B}})\\
\label{eq10}\dot{\mathcal{E}}&=&\frac{\beta}{\alpha}\nabla\times{{\mathcal{B}}-\frac{1}{\gamma}\mathcal{J}}+( \frac{\dot{\beta}}{\beta}{\mathcal{E}})
\end{eqnarray}
 and constraint equations:
 \begin{eqnarray}
\label{eq11} \nabla {.}{\mathcal{B}}&=&0\\
\label{eq12} \nabla{.}{\mathcal{E}}&=&\beta\rho_{0}.
 \end{eqnarray} The constraints are identical to those for the case of constant scalars considered in section (\ref{ST}). Each evolution equation however has an extra term that is modulated by time variations in the pre-factors.

\section{Induction equation}
Substituting Ohm's law into the Faraday's law of induction, and using Ampere's law to eliminate the current density; ${\bold J}$, one can write a single evolution equation for ${\bold B}$, which is called the induction equation:

\begin{eqnarray}
\label{eq17}\dot{\mathcal{B}}&=&\frac{\alpha^2}{\beta^2}\nabla\times(u\times\mathcal{B})-\frac{\alpha\eta}{\gamma\beta}(\nabla\times\mathcal{J})+\frac{\dot{\alpha}}{\alpha}\mathcal{B}\nonumber\\&&~-\frac{\alpha}{\beta}[(u\times\mathcal{B}){.}\nabla(\frac{\alpha}{\beta})+\eta\mathcal{J}{.}\nabla(\frac{1}{\gamma})]\nonumber\\&&~~-\frac{\alpha}{\beta^2}(u\times\mathcal{B}){.}\nabla\beta+\frac{\alpha\eta}{\gamma\beta^2}\mathcal{J}{.}\nabla\beta.
\end{eqnarray}
This the full equation, which does take in account both time and spatially variation on all parameters. Here too we could consider sub cases categorised by whether the variations in the scalar are with respect to time or space.
\subsection{Constant}
Here we consider the case: $\dot{\alpha}=\dot{\beta}=\dot{\gamma}=0$ and 
$\nabla\alpha=\nabla\beta=\nabla\gamma=0,$ which yields:
\begin{eqnarray}
\label{eq18}\dot{\mathcal{B}}&=&\frac{\alpha^2}{\beta^2}\nabla\times(u\times\mathcal{B})-\frac{\alpha\eta}{\gamma\beta}(\nabla\times\mathcal{J}),
\end{eqnarray} which recovers the familiar induction equation.
\subsection{Constant in time}
Here we consider the case
\[\dot{\alpha}=\dot{\beta}=\dot{\gamma}=0\] which then yields:

\begin{eqnarray}
\label{eq20}\dot{\mathcal{B}}&=&\frac{\alpha^2}{\beta^2}\nabla\times(u\times\mathcal{B})-\frac{\alpha\eta}{\gamma\beta}(\nabla\times\mathcal{J})\nonumber\\&&~-\frac{\alpha}{\beta}[(u\times\mathcal{B}){.}\nabla(\frac{\alpha}{\beta})+\eta\mathcal{J}{.}\nabla(\frac{1}{\gamma})]\nonumber\\&&~~-\frac{\alpha}{\beta^2}(u\times\mathcal{B}){.}\nabla\beta+\frac{\alpha\eta}{\gamma\beta^2}\mathcal{J}{.}\nabla\beta
\end{eqnarray} 
This, like the corresponding sub-case of Maxwell equations given above will require gauge choices to be made.

\subsection{Constant in space}
We note from section (\ref{Max}), that one obtains a set of equations which harbour Maxwell's equations as the limiting case. In particular constraint equation in this sub-case satisfy the same conditions as those in the standard Maxwell's equation. We require that \[\nabla\alpha=\nabla\beta=\nabla\gamma=0,\]  from which we obtain the corresponding induction equation:
\begin{eqnarray}
\label{eq19}\dot{\mathcal{B}}&=&\frac{\alpha^2}{\beta^2}\nabla\times(u\times\mathcal{B})-\frac{\alpha\eta}{\gamma\beta}(\nabla\times\mathcal{J})+\frac{\dot{\alpha}}{\alpha}\mathcal{B}.
\end{eqnarray} It is obvious that one will obtain standard Maxwell's induction equation if he or she demands that the scalars be constant with respect to time variations, albeit with some modulation of parameters. Eqn. (\ref{eq19}) suggests that the magnetic field could be induced via dynamics in the media as depicted by the last term on the RHS. We have limited ${\bf\alpha}$ to be a function of location and time only, but  the is no reason why ${\bf\alpha}$ or ${\bf\beta}$ could not be a function of other aspects such as temperature or entropy i.e ${\bf\alpha}={\bf\alpha(r, t, T, S)}$ where ${\bf T}$ is temperature and ${\bf S}$ is entropy respectively.  This expanded scope would allow these scalars to encode thermodynamical properties of the media, and is a subject that will be investigated elsewhere \cite{Bobby}. 

\section{Magnetohydrodynamics}
Magnetohydrodynamics, in the broadest possible sense, could be considered as motion of compressible conducting fluids in the presence of magnetic fields. The motion can be described by kinetic theory together with the laws of thermodynamics. Various natural frequencies of oscillations occur as a result of the electrical and magnetic properties of the fluid. The frequency often considered in this approximation is that of plasma when positive and negative particles are slightly separated and then released. The natural frequency of gyration about the magnetic field is gyro-frequencies of the different charged particles species. 

It is understood that the MHD approximation is applicable if the time variation is much longer that the gyro-period of the heaviest particle species. In this regard, the speeds are much less than the speed of light and standard Maxwell's equation can be simplified as part of low-velocity, low-frequency approximation. 

A fully ionised plasma may be viewed as an assemblage of charged point particles in motion \cite{BoPat,Walk}. In this case, the charge density, {\it q}, and the current density, {\bf J}, both functions of position and time, determine the positions and velocities of such charges. In the standard approximation, the particles are taken as unbound and the magnetic properties arising from the orbital and spin angular momentum are neglected, which then allows the electromagnetic field in such a medium to be described by the two field variables that are appropriately named; {\bf B} (magnetic) and {\bf E} (electric). 

We argue that if appropriately parametrised, e.g $\mathcal {E}$ and $\mathcal{B}$, some of the restrictions mentioned above may be relaxed allowing for wider application of MHD procedures with the added benefit that the standard MHD remains a subset of the results. In this case,
\begin{eqnarray}
\label{eq20}\dot{\mathcal{B}}&=&\frac{\alpha^2}{\beta^2}\nabla\times(u\times\mathcal{B})-\frac{\alpha\eta}{\gamma\beta}(\nabla\times\mathcal{J})+\frac{\dot{\alpha}}{\alpha}\mathcal{B} ,
\end{eqnarray}
will be the relevant induction equation. This could be subjected to the methods in \cite{Bier, Bat}, to look at both magneto-genesis and mean field theory, with the hope understanding  further the theory of magnetic dynamos \cite{Axel, Kulsrud}. This formulation would allow one to investigate magnetic analogues \cite{BoPat}.

 It will be noted that Lorentz \cite{Budd} once showed that if the molecules of a dielectric were arranged in space or if they were arranged in a cubic lattice, then the average electric field acting on the molecule would have two parts; the electric part and the polarisation part. The author of \cite{Budd} suggest that the polarisation can be neglected if one considered that vectors in Maxwell's equations are continuously distributed in space and are constant over distances which are small compared to a wavelength. According to the author, Maxwell's equations therefore imply a 'smoothing-out' process over a sufficiently large distance given the length scales.

 The above argument becomes particularly poignant when one considers the physical requirements for the fluid approximation to be deemed valid. In particular, we know that on the scales of individual particles, the charge and current densities fluctuate wildly, displaying a $\delta$-function behaviour where they appear as zero everywhere except where the particle is located. If a particle is located at position {\bf r} has a charge value {\it q} and velocity {\bf v}, then its charge density $\eta$ may be given by:

\begin{eqnarray}
\eta({\bf r})&=&q\delta^3({\bf r})
\end{eqnarray} and when the charge is found somewhere in a volume {\it V}, 
\begin{eqnarray}
q &=& \int_{V} \eta({\bf r}) dV.
\end{eqnarray} The average charge, denoted by $\langle \eta \rangle$,  in this volume {\it V} is then given by
\begin{eqnarray}
\langle \eta \rangle &=& \frac{1}{V}\int_{V} \eta({\bf r}) dV.
\end{eqnarray} Fluid approximation is valid if the limit approached when {\it V} is made considerably small remains larger than the inter-particle distances involved. One can define the average current density $\langle {\bf J}\rangle$ similarly. Given these length scales, the actual particle distribution may be approximated by the average particle distribution. It is known that a further reduction of the volume to a size comparable to the inter-particle distances invalidate fluid approximation in the standard MHD consideration. If one takes into account kinetic and thermodynamic properties of plasma, then as pointed out in \cite{Bisk}, the conventional theory requires that the mean free path be short compared with typical gradient scales. i.e short in comparison to the length scales in which change is noticeable. While this is usually satisfied for liquids and neutral gases, the mean free path in hot plasmas becomes very long, such that formally the condition for a fluid approximation may easily be violated. This is true even for some astrophysical plasmas of large extent, for instance the solar wind, and much more so for hot laboratory plasmas such as in tokamaks. This implies that the standard setup leaves no room for one to effectively investigate what happens as this limit is approached. We think that the current parametrisation might aid in this regard, a subject that will be considered elsewhere \cite{Bobby}. In particular, it is known that the acceleration of particles at magnetic neutral points in the presence of an electric field produced by plasma convection may in turn generate magnetic field. This represent environmental interaction effect whose consequence was captured by the author of \cite{Dung}, where the author argued that the current produced by motion of  particles would take the form of a thin sheet in which the diffusion of the magnetic field would necessarily dominate. It is noted that the diffusion on the sheet would cause field lines passing through the
it  to change their connectivity to one another.

\section{Discussions and Conclusion}
We have considered a generalisation of Maxwell's equation and extended this to MHD approximation. It is instructive to note that MHD approximation glosses over fundamental physical possibilities. In particular, it is known that plasma behaviour is generally strongly anisotropic, if in the presence of a magnetic field. Consider the case of a collision-less plasma, here the effective mean free path in the direction perpendicular to the field is the gyro-radius which is usually very small. It is also known that in a magnetised plasma gradients parallel to the field where the mean free path is long tend to be much weaker than in the perpendicular direction. 

Secondly, a collision-less plasma experiences dissipation. It is known, for example \cite{Bisk}, that even in the absence of Coulomb collisions, stochastic particle orbits and phasing mixing occur which lead to small-scale plasma turbulence  and hence to efficient dissipation although the velocity distribution functions do not generally relax to a Maxwellian.  In effect, plasma particles generally feel a markedly short effective mean free path, so that for large-scale plasma motions a fluid approximation still holds even in a collision-less plasma.  It is clear that the question of the validity of the ideal fluid approximation calls for a more nuance argument. We know that dissipation effects are represented by various kinds of diffusion processes, which are called weak if the diffusion coefficients is small compared to the spatial scale of interest. It should be noted that the dissipation rate is depended on the local spatial scales and are determined by the internal dynamics. These dynamics could be monitored via the parametrisation we have suggested in this article. I
It is known that the case of nonsingular eigen-modes represent media for which ideal stability theory has a solid foundation, but singular equilibria containing certain discontinuities, such as current sheets, and singular eigen-modes which may be influenced by dissipation effects. Such would require a non-ideal theory. In the case where strongly
non-linear dynamic processes are important, it is known that dissipation is important as large-scale motions may rapidly build up small scale structures. 

We have achieved three things:
1) Introduced an new parametrisation which leads to an expanded set of Maxwell like equations, and which allows for environmental dynamics to be taken into account. 2) We have identified the sub case, from available cases, which specialises the classical Maxwell's equation.
3) We have developed an induction corresponding to case given in the previous step and given the MHD approximation. We have also highlighted areas of potential application.
\section{Acknowledgment}
BO was funded by University of Cape Town's URC grant.

\end{document}